\documentclass[preprint,twocolumn,sort&compress,3p]{elsarticle}
\usepackage{amsmath,amssymb}
\usepackage{graphicx}
\usepackage{slashed}
\usepackage[T1]{fontenc}
\usepackage{color}
\usepackage{hyperref}
\usepackage{times}
\usepackage{overpic}

\newcommand{\e}{\operatorname{e}}
\newcommand{\stext}[1]{\raisebox{-1.5pt}{{\scriptsize #1}}} % for subscripts
\definecolor{tumblau}{cmyk}{1,0.43,0,0}
\definecolor{tumrot}{RGB}{202,033,063}
\definecolor{tumgruen}{cmyk}{0.35,0,1,0.2}

\begin{document}
\title{Functional renormalization group approach to neutron matter}
\author[tum,ect]{Matthias Drews\corref{cor}}
\ead{matthias.drews@ph.tum.de}
\author[tum,ect]{Wolfram Weise}
\ead{weise@tum.de}
\address[tum]{Physik Department, Technische Universit\"{a}t M\"{u}nchen, D-85747 Garching, Germany}
\address[ect]{ECT*, Villa Tambosi, I-38123 Villazzano (Trento), Italy}
\cortext[cor]{Corresponding author}
\date{\today}

\begin{abstract}
The chiral nucleon-meson model, previously applied to systems with equal number of neutrons and protons, is extended to asymmetric nuclear matter. Fluctuations are included in the framework of the functional renormalization group. The equation of state for pure neutron matter is studied and compared to recent advanced many-body calculations. The chiral condensate in neutron matter is computed as a function of baryon density. It is found that, once fluctuations are incorporated, the chiral restoration transition for pure neutron matter is shifted to high densities, much beyond three times the density of normal nuclear matter. 
\end{abstract}
\begin{keyword}
functional renormalization group \sep neutron matter \sep neutron stars
\PACS 26.60.Kp \sep 26.60.-c \sep 21.65.Cd
\end{keyword}

\maketitle

\section{Introduction}
In recent years our understanding of neutron matter has been sharpened significantly. Empirical data as well as theoretical progress set increasingly strong constraints for the equation of state (EoS) at high baryon densities. The observation of two-solar mass neutron stars \cite{demorest2010two-solar-mass,antoniadis2013massive} implies that the EoS must be sufficiently stiff in order to support such dense systems against gravitational collapse. 

At the same time different realistic calculations of neutron matter based on purely hadronic degrees of freedom are seen to be converging to a consistent picture of the energy per particle as a function of neutron density. Approaches such as chiral Fermi liquid theory \cite{holt2013chiral}, chiral effective field theory (ChEFT, \cite{fritsch2005chiral,fiorilla2012chiral,holt2013nuclear}), or quantum Monte Carlo (QMC) calculations \cite{gandolfi2012maximum,gandolfi2014equation} all agree with each other within their ranges of applicability. Whereas compact stars with a considerable ``exotic'' composition, such as a substantial quark core, seem to provide not enough pressure to support a two-solar mass neutron star unless additional strongly repulsive forces are invoked, conventional hadronic matter is consistent with all available mass-radius constraints \cite{hell2014dense}.

In recent publications \cite{drews2014dense,drews2013thermodynamic}, a successful chiral nucleon-meson model for symmetric nuclear matter, previously introduced in \cite{floerchinger2012chemical}, was studied beyond mean-field approximation. Fluctuations were treated within the framework of the functional renormalization group (FRG). The importance of a proper handling of fluctuations around the nuclear liquid-gas phase transition was demonstrated. Moreover, no sign of chiral restoration was found for temperatures below about 100~MeV and densities up to about three times nuclear saturation density, \mbox{$n_0=0.16\text{ fm}^{-3}$}. 

In the present letter we extend this model to asymmetric nuclear matter. The equation of state for pure neutron matter is computed and compared with state-of-the-art many-body calculations. As in symmetric nuclear matter, fluctuations tend to stabilize the hadronic phase characterized by spontaneously broken chiral symmetry and shift the chiral restoration transition to densities much larger than those anticipated in mean-field approximation. This result is of relevance for chiral approaches to strongly interacting, highly compressed baryonic matter, indicating that nucleon and meson (rather than quark) degrees of freedom are still active at densities several times that of normal nuclear matter.

\section{Chiral nucleon-meson model and fluctuations}
The chiral nucleon-meson model is designed to describe nuclear matter and its thermodynamics around the liquid-gas phase transition. The relevant degrees of freedom are protons and neutrons forming an isospin doublet nucleon field \mbox{$\psi=(\psi_p,\psi_n)^T$}. The nucleons are coupled to boson fields: a chiral four-component field $(\sigma,\boldsymbol\pi)$ transforming under the chiral group \mbox{$\operatorname{SO}(4)\cong\operatorname{SU}(2)_L\times\operatorname{SU}(2)_R$}, an isoscalar-vector field $\omega_\mu$ and an isovector-vector field $\boldsymbol\rho_\mu$. Note that these $\omega$ and $\rho$ fields are not to be identified with the known omega and rho mesons. They are introduced here to act as background mean fields representing the effects of short-distance interactions between nucleons, averaged over the baryonic medium. The $\rho$ field appears as an additional degree of freedom in isospin-asymmetric matter, as compared to symmetric nuclear matter where its expectation value vanishes due to isospin symmetry. The Lagrangian of the extended nucleon-meson model reads
\begin{align}\label{eq:Lagrangian}
	\begin{aligned}
		\mathcal L&=\bar\psi i\gamma_\mu\partial^\mu\psi+\frac 12\partial_\mu\sigma\,\partial^\mu\sigma+\frac 12\partial_\mu\boldsymbol\pi\cdot\partial^\mu\boldsymbol\pi \\
		&\quad-\bar\psi\Big[g(\sigma+i\gamma_5\,\boldsymbol\tau\cdot\boldsymbol\pi)+\gamma_\mu(g_\omega\, \omega^\mu+g_\rho\boldsymbol\tau\cdot\boldsymbol\rho^\mu)\Big]\psi \\
		&\quad-\frac 14 F^{(\omega)}_{\mu\nu}F^{(\omega)\mu\nu} - \frac 14 \boldsymbol F^{(\rho)}_{\mu\nu}\cdot\boldsymbol F^{(\rho)\mu\nu} \\
		&\quad+\frac 12m_\omega^2\,\omega_\mu\,\omega^\mu+\frac 12m_\rho^2\,\boldsymbol\rho_\mu\cdot\boldsymbol\rho^\mu- {\cal U}(\sigma,\boldsymbol\pi),
	\end{aligned}
\end{align}
Here $\boldsymbol\tau$ are the isospin Pauli-matrices, and $F_{\mu\nu}^{(\omega)}=\partial_\mu\omega_\nu-\partial_\nu\omega_\mu$, $\boldsymbol F_{\mu\nu}^{(\rho)}=\partial_\mu\boldsymbol\rho_\nu-\partial_\nu\boldsymbol\rho_\mu-g_\rho\,\boldsymbol\rho_\mu\times\boldsymbol\rho_\nu$ (only the three-component in isospin space of the time component of $\boldsymbol\rho_\mu$ will be involved in the further discussions, so the non-abelian part of $\boldsymbol F_{\mu\nu}^{(\rho)}$ is actually not relevant). The potential \mbox{$\mathcal U(\sigma,\boldsymbol\pi)$} has a piece, \mbox{$\mathcal U_0(\chi)$}, that depends only on the chirally invariant square \mbox{$\chi=\frac 12(\sigma^2+\boldsymbol\pi^2)$}, as well as an explicit symmetry breaking term:
\begin{align}
	\mathcal U(\sigma,\boldsymbol\pi)=\mathcal U_0(\chi)-m_\pi^2f_\pi(\sigma-f_\pi)~,
\end{align}
with the pion mass \mbox{$m_\pi=135\text{ MeV}$} and the pion decay constant \mbox{$f_\pi=93\text{ MeV}$}.

As demonstrated in \cite{drews2013thermodynamic}, fluctuations beyond the mean-field approximation can be included using the functional renormalization group approach. A proper treatment of fluctuations turned out to be crucial in order to make contact with results from in-medium chiral perturbation theory calculations of symmetric nuclear matter \cite{fiorilla2012chiral}, emphasizing in particular the role of two-pion exchange dynamics and three-body forces in the nuclear medium. One therefore expects that a full treatment of fluctuations with FRG methods is also important for asymmetric nuclear matter, given the pronounced isospin dependence induced by the fluctuating pion field through multiple pion exchange processes.

The effective action $\Gamma_k$ based on the Lagrangian (\ref{eq:Lagrangian}) depends on a renormalization scale $k$ and interpolates between a microscopic action, \mbox{$\Gamma_{k=\Lambda}$}, defined at an ultraviolet renormalization scale $\Lambda$, and the full quantum effective action, \mbox{$\Gamma_{\stext{eff}}=\Gamma_{k=0}$}. As the scale $k$ is lowered, the renormalization group flow of $\Gamma_k$ is determined by Wetterich's equation~\cite{wetterich1993exact},
\begin{align}\label{eq:Wetterich}
	\begin{aligned}
		k\,\frac{\partial\Gamma_k}{\partial k}=
		\begin{aligned}
			\vspace{1cm}
			\includegraphics[width=0.08\textwidth]{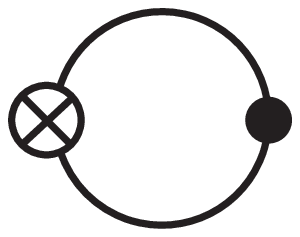}
		\end{aligned} \vspace{-1cm}=\frac 12 \operatorname{Tr}\frac{k\,\frac{\partial R_k}{\partial k}}{\Gamma_k^{(2)}+R_k}\,,
	\end{aligned}
\end{align}
where \mbox{$R_k=(k^2-\boldsymbol p^2)\,\theta(k^2-\boldsymbol p^2)$} is a regulator function and $\Gamma_k^{(2)}=\frac{\delta^2\Gamma_k}{\delta\phi^2}$ is the full inverse propagator. In leading order of the derivative expansion, \mbox{$\Gamma_k= \int d^4x\;\left(\frac 12 \partial_\mu\phi^\dagger\,\partial^\mu\phi+ U_k\right)$}, where $\phi$ symbolizes all appearing fields and $U_k$ is the scale-dependent effective potential. The flow equation reduces now to an equation for $U_k$. In the spirit of Ref.~\cite{litim2006non-perturbative} the flow of the difference
\begin{align}
	\bar U_k(T,\mu_n,\mu_p)=U_k(T,\mu_n,\mu_p)-U_k(0,\mu_c,\mu_c)
\end{align}
is computed, with the effective potential $U_k(T,\mu_n,\mu_p)$ taken at given values of temperature $T$ and of neutron/proton chemical potentials, $\mu_n$ and $\mu_p$, subtracting 
$U_k(0,\mu_c,\mu_c)$ at the liquid-gas transition for symmetric matter at zero temperature. The critical chemical potential $\mu_c=923\text{ MeV}$ at vanishing temperature is the difference between nucleon mass and binding energy. The subtraction at $\mu=\mu_c$ is motivated by the fact that at this point, nuclear physics information can be optimally used to constrain the effective potential. The regime $0\le\mu<\mu_c$ corresponds to a single physical state, the vacuum, with constants $m_\pi$ and $f_\pi$ unchanged by the FRG evolution~\cite{drews2013thermodynamic}. A more detailed discussion will be presented in a forthcoming publication \cite{drews2014}.

The $k$-dependence of $\bar U_k$ is given by the simplified flow equation
\begin{align}
	\begin{aligned}
		{V\over T}\,\frac{k\,\partial\bar U_k}{\partial k}&(T,\mu_n,\mu_p) \\
		&=
		\begin{aligned}
			\hspace{-.1cm}
			\vspace{1cm}
			\includegraphics[width=0.08\textwidth]{wetterich_fermion.eps}
		\end{aligned} \vspace{-1cm}\Bigg|_{T,\mu_n,\mu_p}-
		\begin{aligned}
			\hspace{-.1cm}
			\vspace{1cm}
			\includegraphics[width=0.08\textwidth]{wetterich_fermion.eps}
		\end{aligned} \Bigg|_{\begin{subarray}{l} T=0 \\ \mu_n=\mu_p=\mu_c \end{subarray}}.
	\end{aligned}
\end{align}
The loops symbolize the full propagators of both fermions (nucleons) and bosons (pions and sigma) with inclusion of the regulator.
The heavy vector bosons $\omega_\mu$ and $\boldsymbol\rho_\mu$ are treated as non-fluctuating mean fields. Their Compton wavelengths are supposed to be small compared to the distance scales characteristic of  the Fermi momenta under consideration. Rotational invariance implies that the spatial components of the vector mean fields vanish. The only components that can acquire non-zero expectation values are $\omega_0$ and $\rho_0^3$. Their effect is a shift of neutron and proton chemical potentials according to:
\begin{align}
	\begin{aligned}
		 \mu^{\stext{eff}}_{n,p}&=\mu_{n,p}-g_\omega\,\omega_0\pm g_\rho\,\rho_0^3~.
	\end{aligned}
\end{align}
The scalar boson $\sigma$ and the pions $\boldsymbol\pi$ are light compared to the energy scales we are interested in and so they are allowed to fluctuate. Similarly, the nucleons are kept in the flow equations, thus incorporating soft nucleon-hole excitations around the Fermi surface. Under these conditions, the flow equations for the present model become:
\begin{gather}\label{eq:flow_equation}
	\frac{\partial\bar U_k(T,\mu_n,\mu_p)}{\partial k}=f_k(T,\mu_n,\mu_p)-f_k(0,\mu_c,\mu_c)\,,
\end{gather}
with
\begin{align}
	\begin{aligned}
		&f_k(T,\mu_n,\mu_p)\\
		&=\frac {k^4}{12\pi^2} \bigg\{3\cdot\frac{1+2n_{\stext B}(E_\pi)}{E_\pi}+\frac {1+2n_{\stext B}(E_\sigma)}{E_\sigma} \\
		&\quad-4 \sum_{i=n,p}\frac{1-\sum_{r=\pm1}n_{\stext F}\big(E_{\stext N}-r\mu^{\stext{eff}}_i(k)\big)}{E_{\stext N}}\bigg\}\,.
	\end{aligned}
\end{align}
Here,
\begin{align}
	\begin{gathered}
		E_\pi^2=k^2+U_k'(\chi)\,,\; E_\sigma^2=k^2+U_k'(\chi)+2\chi\, U_k''(\chi)\,, \\
		U_k'(\chi)=\frac{\partial U_k(\chi)}{\partial\chi}\,,\quad E_{\stext N}^2=k^2+2g^2\chi\,, \\
		 \mu^{\stext{eff}}_{n,p}(k)=\mu_{n,p}-g_\omega\,\omega_0(k) \pm g_\rho\,\rho_0^3(k)~,\\
		n_{\stext B}(E)=\frac 1{\e^{E/T}-1}\,,~~\text{ and }~˝\, n_{\stext F}(E)=\frac 1{\e^{E/T}+1}\,.
	\end{gathered}
\end{align}
The $k$-dependent mean fields $\omega_0(k)$ and $\rho_0^3(k)$ are defined at the minima of $U_k$ for each scale $k$. These fields are thus eliminated as external parameters, simplifying the numerical effort. Their values at $k$ are given by the solutions of the following equations which supplement the FRG equation~\eqref{eq:flow_equation}:
\begin{gather}\label{eq:omega_0_rho_0}
	\begin{gathered}
	g_\omega\,\omega_0(k) = \sum_{r=\pm1}\frac {g^2_\omega}{3\pi^2m_\omega^2}\int_k^\Lambda dp \; \frac{p^4}{E_{\stext N}} \hspace{3cm} \\
	\times\frac\partial{\partial \mu}\Big[n_{\stext F}\big(E_{\stext N}-r\mu^{\stext{eff}}_p(k)\big)+n_{\stext F}\big(E_{\stext N}-r\mu^{\stext{eff}}_n(k)\big)\Big] \,, \\
	g_\rho\,\rho^3_0(k) = \sum_{r=\pm1}\frac {g^2_\rho}{3\pi^2m_\rho^2}\int_k^\Lambda dp \; \frac{p^4}{E_{\stext N}} \hspace{3cm} \\
	\times\frac\partial{\partial \mu}\Big[n_{\stext F}\big(E_{\stext N}-r\mu^{\stext{eff}}_p(k)\big)-n_{\stext F}\big(E_{\stext N}-r\mu^{\stext{eff}}_n(k)\big)\Big] \,.
	\end{gathered}
\end{gather}
The ultraviolet potential at \mbox{$k =\Lambda$} is fixed in such a way as to reproduce the mean field potential from Ref.~\cite{drews2013thermodynamic} at \mbox{$T=0$} and \mbox{$\mu_n=\mu_p=\mu_c$}. This guarantees a good description of well-known properties of symmetric nuclear matter around the liquid-gas transition. In fact all parameters apart from $g_\rho$ and $m_\rho$ are determined in this way. The explicit values can be found in Ref.~\cite{drews2013thermodynamic}. With $\rho_0^3$ entering as a mean field, only the ratio $g_\rho^2/m_\rho^2$ appears in the (Hartree type) self-consistent equations. Therefore only a single additional parameter, representing the strength $G_\rho \equiv g_\rho^2/m_\rho^2$ of an equivalent short-distance contact term, $G_\rho (\psi^\dagger \boldsymbol\tau \psi)^2$, is introduced when turning from symmetric to asymmetric nuclear matter and neutron matter. The intermediate and long-range isospin-dependent dynamics is governed entirely by pion degrees of freedom with no additional input required. This renders the model extremely rigid.

\begin{figure}
	\centering
	\begin{overpic}[width=0.45\textwidth]{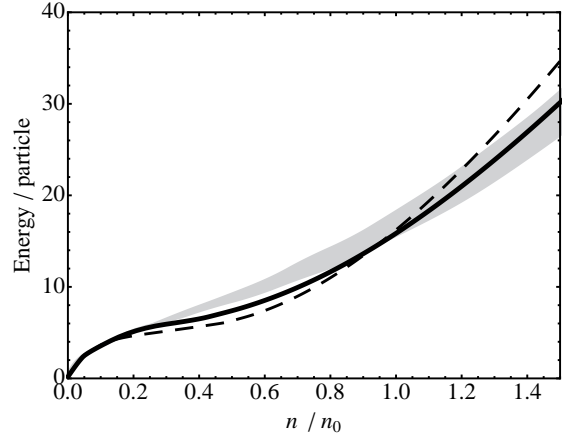}
	\end{overpic}
	\vspace{-0.2cm}
	\caption{Equation of state for small densities for the full FRG calculation (black line) compared with mean field results (dashed line), and chiral Fermi liquid theory~\cite{holt2013chiral} (gray band).\label{fig:eos_low_density}}
\end{figure}

We note that a potential source of isospin breaking is neglected in the present approach.  Introducing an isospin chemical potential, \mbox{$\mu_I = \mu_p - \mu_n$}, the pions are no longer degenerate and $\operatorname{SO}(4)$ is broken to $\operatorname{SO}(2)\times\operatorname{SO}(2)$. As a consequence, the pion field components $\pi_+$ and $\pi_-$ experience the chemical potential $\mu_I$. The complexity of the RG equations increases substantially and the equations for the vector bosons can no longer be integrated since they depend on the potential $U_k$. However the influence of these isospin-breaking terms on the equation of state is expected to be small as pointed out in perturbative calculations 
based on chiral effective field theory \cite{kaiser2012isovector}. All isospin-breaking effects are therefore considered to be absorbed by adjusting the coupling strength $G_\rho$ of the isovector-vector boson.

The full set of equations \eqref{eq:flow_equation} and \eqref{eq:omega_0_rho_0} is solved using the grid method proposed in Ref.~\cite{adams1995solving}. The grand-canonical potential $U_{\stext{gc}}$ is then the effective potential evaluated at its minimum as a function of $\sigma$. From $U_{\stext{gc}}$ all thermodynamic properties can be derived. This framework is now prepared to deal with the thermodynamics
of isospin-asymmetric nuclear matter. In the present paper we focus on the extreme case of pure neutron matter at zero temperature, characterized by vanishing proton density: \mbox{$n_p = -\partial U_{\stext{gc}}/\partial\mu_p=0$}. The general case of non-zero temperatures and varying proton-to-neutron fractions will be investigated in a forthcoming more extended work.

\begin{figure}
	\centering
	\begin{overpic}[width=0.45\textwidth]{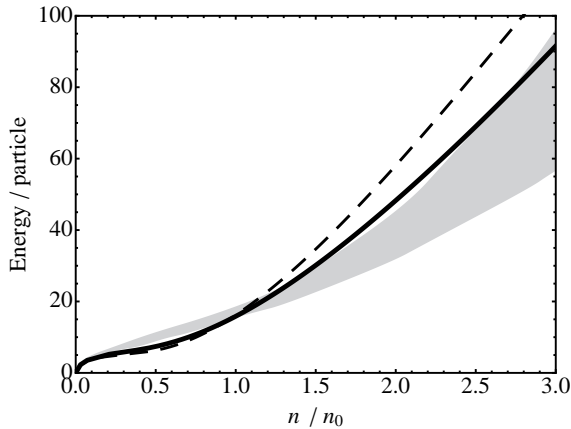}
	\end{overpic}
	\vspace{-0.2cm}
	\caption{Equation of state for the full FRG calculation (black line) compared with mean field results (dashed line), and QMC calculations~\cite{gandolfi2012maximum,gandolfi2014equation} (gray band, with $32.0\text{ MeV}\le E_{\text{sym}}\le33.7\text{ MeV}$).\label{fig:eos_high_density}}
\end{figure}

\section{Results}

The single remaining parameter \mbox{$G_\rho = g_\rho^2/m_\rho^2$} is fixed by reproducing  \mbox{$E_{\stext{sym}}=S(n_0)$}, with the symmetry energy $S(n)$ defined as the difference between the energy per particle of pure neutron matter and symmetric nuclear matter at a given density $n$. Using \mbox{$E_{\stext{sym}}=32~\text{MeV}$}, the value deduced from a large variety of empirical nuclear physics and astrophysics data \cite{lattimer2013constraining}, we find \mbox{$G_\rho =1.12\text{ fm}^2$} (as compared to \mbox{$G_\omega \equiv g_\omega^2/m_\omega^2=4.04\text{ fm}^2$} in the isoscalar sector). 

A second observable of interest is related to the slope of the symmetry energy, $L=3n_0(dS/dn)_{n_0}$. We find a value $L=66.3~\text{MeV}$, reasonably close to the range $L = 40.5\,\text{-}\,61.6\text{ MeV}$ deduced from empirical data and theoretical analysis, see Ref.~\cite{lattimer2013constraining}.

Consider now the energy per particle, $E/N$, of neutron matter. At the mean-field level one encounters the typical problem familiar from relativistic mean field models: $E/N$ comes out too small at low densities, while the EoS is too stiff at large densities compared to realistic many-body calculations. However, the FRG method greatly improves the behavior of the equation of state which is bent towards the band of realistic  many-body calculations as seen in Figs.~\ref{fig:eos_low_density} and \ref{fig:eos_high_density}. There is now much better agreement with results both from chiral Fermi liquid theory \cite{holt2013chiral} at smaller densities and with QMC calculations \cite{gandolfi2012maximum,gandolfi2014equation} up to about three times nuclear saturation density.

\begin{figure}
	\centering
	\begin{overpic}[width=0.47\textwidth]{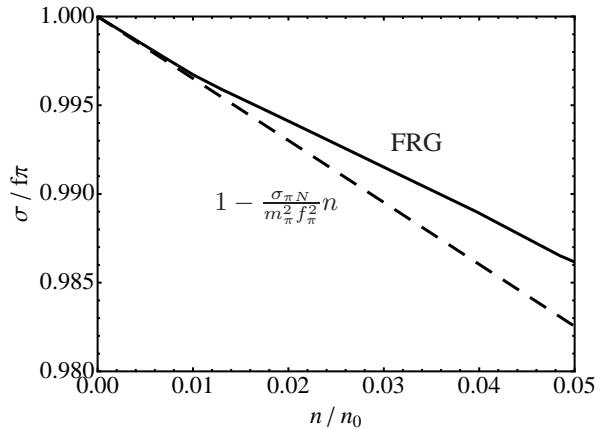}
		\put(35,40){$1-\frac{\sigma_{\pi N}}{m_\pi^2f_\pi^2}n$}
		\put(65,50){FRG}
	\end{overpic}
	\vspace{-0.2cm}
\caption{Chiral order parameter in neutron matter at low densities in comparison with the leading order term using $\sigma_{\pi N}=45\text{ MeV}$.\label{fig:sigma_low_density}}
\end{figure}

\begin{figure}
	\centering
	\begin{overpic}[width=0.47\textwidth]{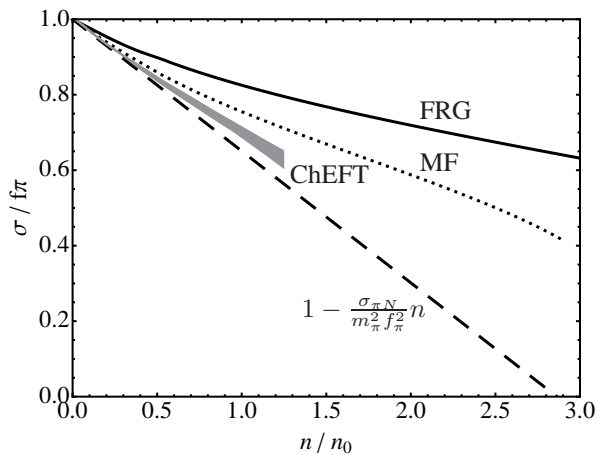}
		\put(70,58){FRG}
		\put(70,49){MF}
		\put(48,47){ChEFT}
		\put(50,24){$1-\frac{\sigma_{\pi N}}{m_\pi^2f_\pi^2}n$}
	\end{overpic}
	\vspace{-0.2cm}
	\caption{Chiral order parameter in neutron matter: linear approximation (dashed line), the ChEFT result \cite{krueger2013chiral} (gray band), the mean-field approximation of the present model (MF - dotted curve) and the full FRG result (solid curve). \label{fig:sigma_high_density}}
\end{figure}

Another interesting issue is the question of chiral symmetry restoration in dense neutron matter. An order parameter of spontaneously broken chiral symmetry is the chiral (quark) condensate, $\left\langle\bar qq\right\rangle$. To leading order in the density $n$,
\begin{align}
	\frac{\left\langle\bar qq\right\rangle_n}{\left\langle\bar qq\right\rangle_0}
=1-\frac{\sigma_{\pi N}}{f_\pi^2m_\pi^2}n\,,
\end{align}
where the slope is determined by the pion-nucleon sigma term, \mbox{$\sigma_{\pi N}=45\pm 5\text{ MeV}$} \cite{gasser1991sigma-term}. In the present model the expectation value of the $\sigma$ field is directly related (proportional) to the chiral condensate. A fit to the full FRG result at low densities gives \mbox{$\sigma_{\pi N}=44\text{ MeV}$} as shown in Fig.\,\ref{fig:sigma_low_density}. Already at about \mbox{$0.01~n_0$}, corresponding to a Fermi momentum slightly less than $m_\pi/2$, the onset of a deviation from the linear behavior is observed. The non-linear structure of the effective potential as a function of the chiral field \mbox{$\chi=\frac 12(\sigma^2+\boldsymbol\pi^2)$} generates many-body forces involving three and more nucleons. Their effects become increasingly important as the density rises, inducing a significant deviation from the linear dropping of the in-medium chiral condensate as a function of $n$. This non-linear behavior appears already at mean-field level and is further enhanced when fluctuations are incorporated in the full FRG result (see Fig.~\ref{fig:sigma_high_density}). As a consequence the chiral condensate is stabilized and the system remains in the hadronic (Nambu-Goldstone) phase with spontaneously broken chiral symmetry up to densities much larger than three times $n_0$. 

Pushing the model to its limits, one observes a rapid crossover to chiral symmetry restoration at about seven to eight times nuclear saturation density once fluctuations are included. In the mean-field approximation, the transition is first order instead and appears at lower density. Similar conclusions have
been drawn in studies using the linear sigma model. Once loop corrections are included, the parameters are re-adjusted in order to avoid unphysically large corrections to three-body forces \cite{nyman1976abnormal,nyman1976chiral,nyman1977abnormal}. One finds that the phase-transition into the abnormal Lee-Wick phase \cite{lee1974vacuum} sets in only at very large densities. Although the models differ, as does the treatment of fluctuations (which is fully non-perturbative in our present approach), the results of both calculations underline the important stabilizing properties of fluctuations.

A perturbative calculation of the quark condensate in neutron matter using chiral effective field theory at next-to-next-to-next-to leading order (N${}^3$LO) \cite{krueger2013chiral}, applicable up to about \mbox{$n \simeq n_0$}, is shown for comparison in Fig.\,\ref{fig:sigma_high_density}. This calculation also features a delay in the tendency towards chiral restoration, but less prominently so than the present FRG calculation. The stability of $\langle\bar{q}q\rangle$ even in highly compressed, cold matter suggests that a chiral approach based on baryon and meson (rather than quark and gluon) degrees of freedom can presumably be extended to quite high densities. In our FRG approach the crossover to chiral symmetry restoration would set in at about six times nuclear saturation density. The resulting neutron matter equation of state, taken as an input for solving the Tolman-Oppenheimer-Volkoff equation, meets all mass-radius constraints from neutron star observations, as will be demonstrated in a forthcoming publication. 

\section{Summary and conclusions}
A chiral nucleon-meson model extended to asymmetric nuclear matter has been studied for the first time with systematic inclusion of fluctuations beyond mean-field approximation, using the framework of the functional renormalization group. Further to the previous treatment of symmetric nuclear matter and its thermodynamics, a single additional parameter, representing the isospin-dependent part of the short-distance nucleon-nucleon interaction, has been fitted to the symmetry energy. Isospin-dependent dynamics at intermediate and long distances is completely determined by multiple pion exchange mechanisms generated non-perturbatively by the renormalization group equations. The resulting equation of state for neutron matter is in good agreement with advanced many-body calculations over a large density range. Chiral symmetry restoration in cold neutron matter is found to be shifted to high densities, considerably beyond at least three times the density of normal nuclear matter.

\section*{Acknowledgments}
This work is supported in part by BMBF and by the DFG Cluster of Excellence ``Origin and Structure of the Universe.''

\bibliographystyle{physlett}
\bibliography{biblio}

\end{document}